\def\papertitle{Differentiable Articulatory Copy-Synthesis of Biphonic Singing}
\def\paperauthorA{Mateo Cámara}
\def\paperauthorB{María Pilar Daza-Llin}
\def\paperauthorC{Fernando Marcos-Macías}
\def\paperauthorD{José Luis Blanco}

\documentclass[twoside,a4paper]{article}
\usepackage{etoolbox}

\usepackage[print]{dafx26v3}

\usepackage{amsmath,amssymb,amsfonts,amsthm}
\usepackage{siunitx}
\usepackage{euscript}
\usepackage[T1]{fontenc}
\usepackage[utf8]{inputenc}
\usepackage{ifpdf}
\usepackage[english]{babel}
\usepackage{caption}
\usepackage{subfig} 
\usepackage{color}
\usepackage{booktabs}
\usepackage{lipsum}
\usepackage{multirow}
\usepackage{soul}

\input glyphtounicode
\ninept

\newcounter{numauth}
\newcounter{listcnt}
\newcommand\authcnt[1]{\ifdefined#1 \stepcounter{numauth} \fi}

\newcommand\addauth[1]{
\ifdefined#1 
\stepcounter{listcnt}
\ifnum \value{listcnt}<\value{numauth}
\appto\authorslist{, #1}
\else
\appto\authorslist{~and~#1}
\fi
\fi}
\authcnt{\paperauthorB}
\authcnt{\paperauthorC}
\authcnt{\paperauthorD}
\authcnt{\paperauthorE}
\authcnt{\paperauthorF}
\authcnt{\paperauthorG}
\authcnt{\paperauthorH}
\authcnt{\paperauthorI}
\authcnt{\paperauthorJ}
\def\authorslist{\paperauthorA}
\addauth{\paperauthorB}
\addauth{\paperauthorC}
\addauth{\paperauthorD}
\addauth{\paperauthorE}
\addauth{\paperauthorF}
\addauth{\paperauthorG}
\addauth{\paperauthorH}
\addauth{\paperauthorI}
\addauth{\paperauthorJ}

\usepackage{times}

\newif\ifpdf
\ifx\pdfoutput\relax
\else
   \ifcase\pdfoutput
      \pdffalse
   \else
      \pdftrue
   \fi
\fi

\ifpdf 
  \usepackage[pdftex,
    pdftitle={\papertitle},
    pdfauthor={\authorslist},
    pdfsubject={Proceedings of the 29th International Conference on Digital Audio Effects (DAFx26)},
    colorlinks=false, 
    bookmarksnumbered, 
    pdfstartview=XYZ 
  ]{hyperref}
  \usepackage[pdftex]{graphicx}
\else 
  \usepackage[dvips]{epsfig,graphicx}
  \usepackage[dvips,
    pdftitle={\papertitle},
    pdfauthor={\authorslist},
    pdfsubject={Proceedings of the 29th International Conference on Digital Audio Effects (DAFx26)},
    colorlinks=false, 
    bookmarksnumbered, 
    pdfstartview=XYZ 
  ]{hyperref}
\fi
\usepackage[hypcap=true]{caption}
\title{\papertitle}

\affiliation
{\paperauthorA, \paperauthorB, \paperauthorC, and \paperauthorD}
{Signal Processing Applications Group \\ Information Processing and Telecommunications Centre \\ Universidad Politécnica de Madrid, Spain\\
{\tt \href{mailto:mateo.camara@upm.es}{mateo.camara@upm.es}}
}

\begin{document}
\ifpdf 
  \DeclareGraphicsExtensions{.png,.jpg,.pdf}
\else  
  \DeclareGraphicsExtensions{.eps}
\fi


\maketitle

\begin{abstract}
Sygyt is a Tuvan style of biphonic singing in which a low vocal drone is sustained while a high harmonic is selectively amplified in the 1--3\,kHz region. Copy-synthesizing this effect remains challenging for articulatory models, since it requires fine control of narrowly focused resonances that standard low-dimensional tract parameterizations cannot easily reproduce. We address this problem with a differentiable Kelly--Lochbaum waveguide augmented with a sublingual second source, cubic B-spline tract parameterization, and spatially varying learnable damping, optimized end-to-end by gradient descent from audio. On 20 segments from two independent sygyt datasets (5 singers, 10 pitches), the proposed model reduces log-spectral distance by 30--38\% relative to an articulatory baseline, with the largest gains concentrated in the overtone region. Cepstral-envelope analysis further shows more accurate recovery of the merged formant structure characteristic of sygyt production. The model also outperforms a DDSP harmonic-plus-noise baseline with direct per-harmonic spectral control, suggesting that explicit acoustic structure is a useful inductive bias for overtone-singing copy-synthesis.
\end{abstract}

\section{Introduction}

\begin{figure}[th]
\centering
\includegraphics[width=\columnwidth]{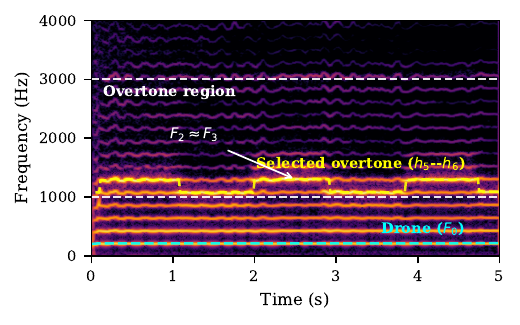}
\caption{Spectrogram of a \emph{sygyt} performance: $h_5$--$h_6$ overtone selection. The singer sustains a drone at $f_0$ (cyan) while shaping the vocal tract to selectively amplify a single overtone (yellow). \textcolor{black}{The 1--3\,kHz overtone region is outlined in white.}}
\label{fig:what_is_sygyt}
\end{figure}

Biphonic singing, known as \emph{khoomei} in Tuva, is a vocal technique that enables a single performer to produce two simultaneous percepts: a sustained fundamental drone and a selectable melodic overtone shaped through vocal tract control. Distinct styles include \emph{sygyt} (high-pitched whistling overtones), \emph{kargyraa} (low-register nonlinearities), and others. This paper focuses on \emph{sygyt}, which is especially associated with narrow-band spectral focusing, often described as formant merging, in the 1--3\,kHz region~\cite{bergevin2020}. Fig.~\ref{fig:what_is_sygyt} illustrates this behavior, showing a prominent narrowband energy ridge superimposed on the fundamental drone.

Conventional synthesis approaches to overtone singing have relied on manually tuned formant synthesis~\cite{kob2002}, coupled-oscillator models~\cite{sakakibara2001,sakakibara2002}, and source–filter decomposition~\cite{tsai2014}. These methods yielded valuable insights but depend heavily on handcrafted control signals, expert analysis, or style-specific assumptions. Meanwhile, articulatory models based on Kelly–Lochbaum waveguides offer a physically meaningful representation of voice production, only recently have such models been optimized inversely for copy-synthesis from recorded speech and singing~\cite{sudholt2023vocal,camara2023optimization}. A key question that remains open is whether gradient-based optimization of a differentiable articulatory model can recover tract shapes that both reproduce recorded \emph{sygyt} acoustics and align with the established formant-focusing mechanism.

\sloppy
We address this question using differential models to produce high-quality synthetic records to match the originals (\textcolor{black}{Articulatory Copy Synthesis}). We confront differential digital signal processing (DDSP) and an extended differentiable articulatory Kelly–Lochbaum model~\cite{kelly1962}. The proposed formulation introduces a sublingual secondary source, a cubic B-spline tract parameterization, spatially varying learnable damping, and an overtone-salience loss that emphasizes harmonic consistency in the target overtone region. The goal is to fit \emph{sygyt} spectra and to determine a physically grounded model that matches acoustic features while maintaining interpretable parameters.

\sloppy
We evaluate the methods on 20 studio-recorded \emph{sygyt} segments drawn from two independent datasets (5 singers, 10 pitches), comparing a Pink Trombone-style articulator-chain baseline~\cite{camara2023optimization}, the proposed B-spline formulation, and a DDSP baseline. The proposed model improves substantially over the articulatory baseline and yields recovered tract configurations that are acoustically interpretable and qualitatively consistent with formant-focusing behavior in \emph{sygyt}. Additional experiments on independent recordings suggest robustness across performers and pitches.

The main contributions of this work are as follows:
\begin{itemize}
\item We formulate \emph{sygyt} copy-synthesis from recorded audio as a differentiable optimization problem. 
\item We introduce an extended Kelly-Lochbaum model that features a sublingual secondary source, smooth tract control, varying damping, and an overtone-aware objective.
\item We demonstrate, on two independent datasets, that this formulation outperforms a standard articulatory baseline and matches the performance of a DDSP-based approach, with supporting subjective preference tests.

\item We show that the optimized solutions reproduce the \textcolor{black}{spectro-temporal} signature of \emph{sygyt}. Ablation results indicate that the sublingual source is the dominant contributor. Subjective tests and objective metrics support our results.
\end{itemize}

\section{Related Work}\label{sec:related}

\begin{figure}[!b]
    \centering
    \begin{minipage}[c]{0.48\columnwidth}
        \centering
        \includegraphics[width=\textwidth]{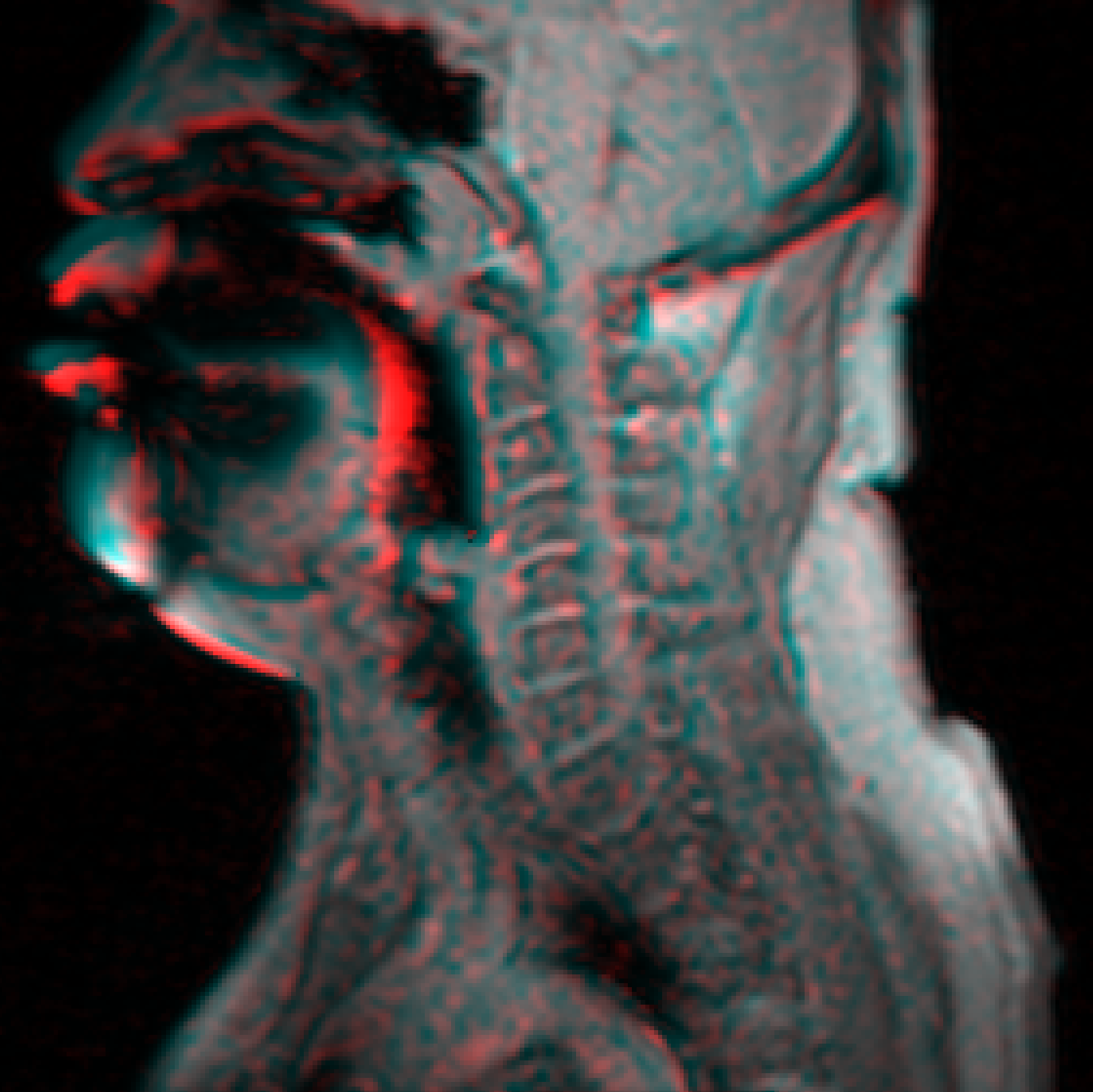}
    \end{minipage}
    \hfill
    \begin{minipage}[c]{0.48\columnwidth}
        \centering
        \includegraphics[width=\textwidth]{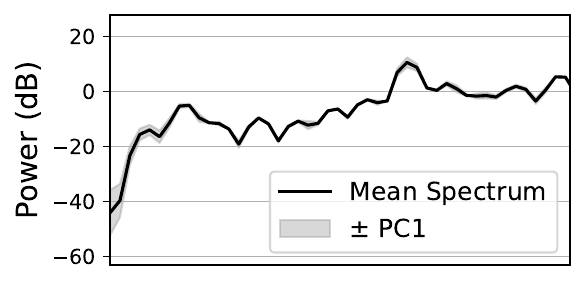}
        \\[0.2cm] 
        \includegraphics[width=\textwidth]{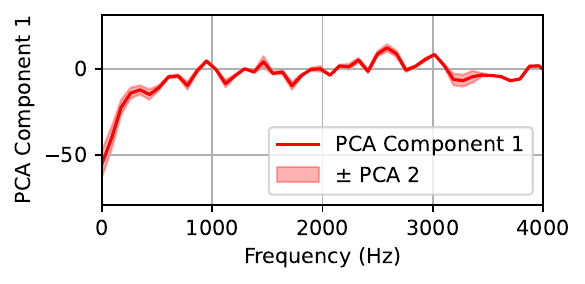}
    \end{minipage}
    
    \caption{Visualizing articulatory and acoustic dynamics in \emph{sygyt}~\cite{bergevin2020}. \textbf{Left:} Average MRI frames (gray) overlaid with PC1 (red) to capture dominant temporal variations beyond the mean variance. \textbf{Right:} Spectrograms featuring LTAS (black) and acoustic PCA (red) displayed above for spectral comparison.}
    \label{fig:mri-mean-pca}
\end{figure}

Magnetic resonance imaging (MRI) has played a central role in revealing the articulatory configurations underlying the highly selective resonances of overtone singing. Fig.~\ref{fig:mri-mean-pca} provides evidence showing synchronized midsagittal MRI data (left) alongside acoustic analysis (right) for a recorded segment of \emph{sygyt}~\cite{bergevin2020}. These measurements provide the link between vocal-tract shapes and the spectra for the sounds that characterize this singing style.

Earlier studies sought to formalize these relationships between articulation and acoustics. For instance, Laver~\cite{laver1991} related vocal-tract configurations to long-term average spectra (LTAS) to mitigate noise effects and spurious variations. Subsequent work employed principal component analysis (PCA) to link spectral deviations with articulatory and resonant variations~\cite{Blanco2013}. Such approaches are particularly revealing for overtone singing styles like \emph{sygyt}, where the acoustic output depends on highly constrained resonant structures and fine posterior tongue adjustments. However, these studies remain primarily descriptive, lacking the controllable or differentiable synthesis frameworks required for direct inverse parameter estimation.


\subsection{Acoustic Analysis of Biphonic Singing}
\label{subsec:formant}

Early acoustic studies sought to explain how overtone singing isolates and amplifies individual harmonics. Bloothooft \textit{et al.}~\cite{bloothooft1992} interpreted the overtone as arising from closely spaced formants, while Klingholz~\cite{klingholz1993} reported substantial formant bandwidth reduction, with resonances narrowing to approximately 20\,Hz under pharyngeal constriction. Adachi and Yamada~\cite{adachi1999} further showed that \emph{sygyt} production can be explained by the convergence of the second and third formants ($F_2 \approx F_3$), thereby creating a narrow spectral peak that selectively reinforces one harmonic component.

Subsequent work provided a more detailed articulatory interpretation of these acoustic observations. Kob~\cite{kob2004} used impedance analysis to show that large overtone amplification can arise from the coincidence of resonances, while Bergevin \textit{et al.}~\cite{bergevin2020}, combining dynamic MRI and studio-quality recordings, argued that \emph{sygyt} can largely be explained by linear filtering, that is, by precise vocal-tract shaping rather than by an additional nonlinear sound source. Sundberg \textit{et al.}~\cite{sundberg2023} and Hefele \textit{et al.}~\cite{hefele2019} further documented the relevant vocal-tract configurations and their acoustic consequences. These studies indicate that overtone singing, and \emph{sygyt} in particular, relies on sharp resonant focusing and strong spectral sensitivity to small articulatory changes.

\subsection{Synthesis of Overtone Singing}

Several synthesis approaches have been proposed for overtone singing, although they differ substantially in their degree of physical grounding and parameter interpretability. Kob~\cite{kob2002} achieved formant-based synthesis of \emph{khoomei} by manually specifying formant frequencies and bandwidths, obtaining perceptually convincing results at the cost of expert parameter tuning. Tsai \textit{et al.}~\cite{tsai2014} transformed conventional singing into overtone singing through source--filter decomposition and spectral-envelope modification, but without recovering an explicit vocal-tract configuration.

Articulatory synthesis offers a more interpretable alternative; control variables correspond directly to vocal production. Classical articulatory models~\cite{sondhi1985,story2013, mullen2004acoustical, murphy2015articulatory} simulate realistic vocal-tract acoustics and have been successfully used in speech production studies. Story's TubeTalkerPkg~\cite{story2013}, for instance, was employed by Bergevin \textit{et al.}~\cite{bergevin2020} to study formant focusing through forward synthesis from prescribed tube-area functions. Similarly, the Pink Trombone real-time synthesizer~\cite{thayer2017} implements an interactive Kelly--Lochbaum-style waveguide model. These systems are primarily designed for forward control and exploration rather than inverse estimation. In overtone singing, particularly in \emph{sygyt}, small changes in tract geometry can produce changes in overtone placement and bandwidth~\cite{bergevin2020, klingholz1993,adachi1999}, degrading our results.

\subsection{Differentiable Sound Synthesis}

Recent work in differentiable audio synthesis~\cite{engel2020,hayes2023} has shown that gradient-based optimization can be combined with structured synthesis models to perform copy-synthesis while preserving meaningful control parameters. Most of this literature has focused on neural synthesizers, differentiable signal processors, or hybrid physical--neural architectures. By contrast, articulatory waveguide models expose parameters with direct physical interpretation, such as cross-sectional areas, damping coefficients, and glottal source characteristics. 

Applying differentiable optimization to overtone singing remains challenging because of the strongly underdetermined inverse mapping from sound to articulation. The target acoustics depend on narrow, high-Q resonances that are highly sensitive. Prior articulatory models have supported forward analysis and formant-targeted synthesis, but we found no evidence of gradient-based copy-synthesis of recorded overtone singing using a differentiable articulatory model with physically interpretable parameters. Likewise, we found no prior computational study directly testing whether gradient-based optimization can recover the formant-merging configurations reported in acoustic analyses of \emph{sygyt}~\cite{adachi1999,bergevin2020}. This gap motivates the present work.

\section{Method}\label{sec:method}

This section describes the proposed \emph{sygyt} copy-synthesis, including the DDSP baseline and the physically-informed differentiable waveguide architecture. 

\subsection{DDSP baseline}\label{sec:baselines}

As a non-physical reference with direct spectral control, we include a DDSP-style synthesizer~\cite{engel2020}. The model generates audio from multiple oscillators. Starting on the primary $f_0$ track, we sum 60 sinusoidal harmonics with learnable amplitudes, plus filtered noise from a 257-bin linear-phase FIR filter (Hann window,  FFT size 512, 50\% overlap). Each harmonic amplitude is independently optimized by gradient descent with respect to the target recording, thereby giving the model direct control over the energy at each frequency. Each frame has 318 free parameters (60 harmonic amplitudes $+$ 1 overall amplitude $+$ 257 noise-filter coefficients). The DDSP synthesizer operates in signal space rather than physical space. This baseline provides a complementary perspective to the psychologically-informed, which must discover spectral structure through tract geometry.

\subsection{Differentiable source-filter model}

\textbf{Extended waveguide}. Our synthesizer extends the differentiable model and efficient JAX implementation of VocalTrax~\cite{mo2024articulatory}. We included three architectural additions: a nasal tract, a sublingual second glottal source, and learnable spatially varying damping. The vocal tract is discretized into three coupled waveguide sections for the \textbf{oral tract}--44 cylindrical segments from glottis to lips (following~\cite{mo2024articulatory})-- the \textbf{nasal tract} --28 segments from velum to nostrils--, and the \textbf{sublingual tube}-- 15 segments connecting the secondary source to the oral tract\textcolor{black}{; segment counts follow the discretization in~\cite{mo2024articulatory} and the simulation runs at 16\,kHz}.

Two \textbf{three-way junctions} couple the waveguide sections. At the sublingual junction (oral segment~9), three branches meet: the oral tract splits into a proximal section (segments 0--8) and a distal section (segments 10--43), connected to the sublingual tube. At the velum junction (segment~17), the oral and nasal tracts are coupled.

\noindent\textbf{Propagation}. Wave propagation follows the standard Kelly-Lochbaum scattering equations. At each segment boundary~$i$, leftward~($L$) and rightward~($R$) traveling waves interact via reflection coefficient $r_i = (A_i - A_{i+1}) / (A_i + A_{i+1})$, where $A_i$ is the cross-sectional area of segment~$i$. In the standard Kelly-Lochbaum model, damping is applied uniformly across all segments. We extend this to a \textbf{learnable damping} $d_i \in [0.99, 0.9999]$ for the oral and sublingual cavity: $R_i \leftarrow R_i \cdot d_i, \quad L_i \leftarrow L_i \cdot d_i$. 

This provides fine-grained Q-factor control: values closer to unity ($d \to 0.9999$) preserve wave energy and sustain sharper resonances (higher Q-factor), while lower values ($d \to 0.99$) increase energy dissipation and broaden formant bandwidth (lower Q-factor). The optimizer can thus selectively sharpen formants near the overtone frequency while attenuating competing resonances elsewhere. Physically, $d_i$ models frequency-dependent losses from viscous boundary layers, heat conduction, and soft tissue absorption, which vary along the vocal tract due to differences in tissue composition, surface area, and constriction geometry~\cite{sondhi1985}. 

The scattering at each three-way junction depends on the areas of the three connecting segments, $A_L$, $A_R$, $A_B$:
\begin{equation}
  \Sigma = A_L + A_R + A_B, \quad
  r_X = \frac{2A_X - \Sigma}{\Sigma}, \quad X \in \{L, R, B\}
\end{equation}


\noindent\textbf{Sources: tone and overtone}.  The glottal and sublingual sources provide the excitations to the waveguide. The latter focuses on the overtone frequency $f_\text{ot}$, injected into the vocal tract via the sublingual junction. This models the hypothesis that diphonic singing involves a secondary vibrating structure below the tongue~\cite{sakakibara2001}.

Sources use a Liljencrants-Fant (LF) waveform with learnable parameters: \textbf{amplitude} $a \in [0.1, 1.0]$--relative to primary source--, \textbf{tenseness} $\tau \in [0.1, 1.0]$-- controls spectral slope--, \textbf{open quotient offset} $\Delta\text{OQ} \in [-0.15, +0.15]$--, \textbf{spectral tilt offset} $\Delta\text{tilt} \in [-2, +2]$.

The secondary $f_{ot}$ is extracted from the target audio as described in Section~\ref{sec:f0_detection} and held fixed during optimization. Crucially, the sublingual tube introduces a \emph{second resonant cavity} parallel to the oral tract, whose interaction with the main tract at the three-way junction produces additional spectral features beyond what a single-tube model can achieve.

\subsection{Tract Parameterizations}\label{sec:parameterization}

We compare two strategies for controlling the vocal tract. DOFs stand for Degrees of Freedom assigned in the models involving the parameters for the waveguide:

\noindent\textbf{Articulator chain} ($\sim$13 DOFs/frame). Following the classical configuration for the Kelly-Lochbaum~\cite{camara2023optimization}, the tract shape is controlled through a list of physiologically-informed articulators: tongue position and diameter (2), throat constriction (1), lip rounding (1), velum opening (1), sublingual coupling (1), and global damping (1). This parameterization guarantees human-like shapes by construction but limits the optimizer's ability to form the narrow resonances required for overtone focusing. The nasal waveguide is not parameterized but included in the model to incorporate nasalization effects and velum movement. 

\noindent\textbf{Cubic B-spline spatial basis} ($\sim$70 DOFs/frame). Unconstrained per-segment optimization produces physically implausible discontinuities. Following an MRI study of vocal-tract geometry during overtone singing, we represent both the oral and sublingual diameter profiles, as well as their damping profiles, using separate sets of cubic B-spline control points placed at anatomically motivated landmarks \textcolor{black}{(glottis, sublingual junction, velum, and lips)}. While the cubic basis increases the number of coefficients, it ensures $C^2$ continuity and remains physically plausible.

Let $\boldsymbol{\alpha}_A \in \mathbb{R}^{K_A}$ and $\boldsymbol{\alpha}_d \in \mathbb{R}^{K_d}$ be the diameter and damping coefficients respectively, where $K=20$ (empirically selected attending to our analysis of the implementation in~\cite{mo2024articulatory} ) are the B-spline control points \textcolor{black}{placed at uniform knot spacing}. Segment-wise profiles are obtained by evaluating the B-spline bases at segment centers:
\begin{equation}
  \mathbf{A} = B_A \boldsymbol{\alpha}_A,
  \qquad
  \mathbf{d} = \mathrm{clip}(B_d \boldsymbol{\alpha}_d,\; 0.99,\; 0.9999),
\end{equation}
where $B_A$ and $B_d$ are the basis matrices \textcolor{black}{(cubic B-spline bases of size $N\times K_A$ and $N\times K_d$ evaluated at the $N$ segment centers)}. Area and damping are applied locally during synthesis but controlled by smooth B-spline parameterizations rather than independent per-segment variables.

Both parameterizations share identical primary and secondary source models (5 each), as well as the same loss functions and optimizer settings. The only difference is how the vocal tract shape and damping are parameterized.

\subsection{Computation}\label{sec:f0_detection}

\noindent\textbf{Tones extraction}. Both the primary fundamental frequency $f_0$ and the overtone frequency $f_\text{ot}$ are extracted from the target audio in a preprocessing step shared by all synthesis conditions. The primary $f_0$ is estimated using the YIN algorithm~\cite{yin2002}.

The overtone frequency is identified via spectral peak analysis on a high-resolution STFT ($n_\text{fft}=4096$). For each voiced frame, harmonics $H_2$--$H_{10}$ of the primary $f_0$ are examined: the \emph{enhancement} of harmonic $H_k$ is defined as the difference between its measured energy and the energy expected under a natural spectral rolloff of 6\,dB per octave. A harmonic is flagged as the active overtone when its enhancement exceeds 6\,dB\textcolor{black}{; if several harmonics exceed the threshold, the one with the largest enhancement is selected}.

To ensure temporal coherence, two mechanisms prevent erratic harmonic switching: (1)~hysteresis requiring $>$3\,dB improvement before changing the active harmonic, and (2)~Savitzky-Golay smoothing (window\,$=$\,11, order\,$=$\,2) of the harmonic-number track. Both $f_0$ and $f_\text{ot}$ trajectories are held fixed during optimization — the synthesis methods differ only in how they use them.


\noindent\textbf{Optimization}. Copy-synthesis optimizes tract shape and source parameters to minimize the reconstruction loss:
\begin{equation}
  \mathcal{L} = \mathcal{L}_\text{STFT} + \lambda_\text{mel}\mathcal{L}_\text{mel}
  + \lambda_\text{harm}\mathcal{L}_\text{harm}
  + \lambda_\text{energy}\mathcal{L}_\text{energy}
  + \lambda_\text{ot}\mathcal{L}_\text{ot}
\end{equation}
\textcolor{black}{In our experiments, we set all loss weights to unity ($\lambda_\text{mel} = \lambda_\text{harm} = \lambda_\text{energy} = \lambda_\text{ot} = 1.0$)}

The multi-resolution STFT loss $\mathcal{L}_\text{STFT}$ uses FFT sizes \{512, 1024, 2048\} with both spectral convergence (Frobenius norm) and log-magnitude (L1) terms~\cite{engel2020}. The mel loss $\mathcal{L}_\text{mel}$ operates on 80-band mel spectrograms, providing perceptually-weighted frequency resolution. The harmonic energy loss $\mathcal{L}_\text{harm}$ matches energy at integer multiples of $f_0$ to preserve the harmonic structure critical for overtone singing. The overtone salience loss $\mathcal{L}_\text{ot}$ penalizes deviation between the synthesized and target overtone:
\begin{equation}
  S_\text{ot}(t) = 10\log_{10}\frac{\sum_{f \in B_h} |X(f,t)|^2}
  {\sum_{f \in B_\pm} |X(f,t)|^2 + \epsilon}
\end{equation}
where $B_h$ is a frequency band \textcolor{black}{of half-width 50\,Hz} centered on the target harmonic $h(t) \cdot f_0(t)$\textcolor{black}{, $h(t) \in \{2,\dots,10\}$ is the active overtone index,} and $B_\pm$ covers \textcolor{black}{the two adjacent harmonics at $(h{\pm}1)\,f_0(t)$ with the same 50\,Hz half-width}\textcolor{black}{; $\epsilon = 10^{-10}$ avoids division by zero}. \textcolor{black}{The overtone-salience loss is then $\mathcal{L}_\text{ot} = \| S_\text{ot}^\text{target} - S_\text{ot}^\text{synth}\|_2^2$.}

All conditions are optimized with Adam using a learning rate of $\eta = 5 \times 10^{-3}$, preceded by a 100-step warm-up from $5 \times 10^{-4}$, gradient clipping at a maximum norm of 1.0, and a 500 iterations budget. The warm-up stage is important for reducing early instability caused by strongly nonlinear waveguide dynamics. The full pipeline is differentiable via JAX, enabling gradient computation for the sample-by-sample waveguide simulation, scattering junctions, and LF source generation. All experiments are run on CPU (AMD Ryzen 7), requiring approximately 30 minutes for a 500-iteration optimization at 16\,kHz. 
Table~\ref{tab:comparison_setup} summarizes the shared experimental protocol.


\begin{table}[!h]
\centering
\caption{Experimental protocol. Components above the mid-rule are shared
across all conditions; components below differ.}
\label{tab:comparison_setup}
\resizebox{\columnwidth}{!}{%
\begin{tabular}{@{}l c c@{}}
\toprule
Component & Waveguide & DDSP \\
\midrule
Target input & \multicolumn{2}{c}{Audio recording} \\
$f_0$ extraction & \multicolumn{2}{c}{YIN + overtone det.\ (fixed)} \\
Optimizer & \multicolumn{2}{c}{Adam ($\eta\!=\!5\!\times\!10^{-3}$, warmup)} \\
Iterations & \multicolumn{2}{c}{500} \\
$\mathcal{L}_\text{STFT}\!+\!\mathcal{L}_\text{mel}\!+\!\mathcal{L}_\text{energy}$ & \multicolumn{2}{c}{\checkmark} \\
\midrule
Synthesis engine & Kelly-Lochbaum waveguide & Harmonic + noise \\
Parameterization & Articulator / B-spline & 60 harm.\ amps + 257-bin FIR \\
Pitch structure & Dual ($f_0 + f_\text{ot}$) & Single ($f_0$) \\
$\mathcal{L}_\text{ot}$ (overtone) & \checkmark & --- \\
\bottomrule
\end{tabular}}
\end{table}


\section{Experimental results}\label{sec:results}

\subsection{Dataset and Setup}

We evaluate copy-synthesis on 20 recordings of Tuvan \emph{sygyt} (overtone singing) drawn from two independent studio-recorded datasets. The first comprises 10 recordings from the HFA Overtone Singing Preview dataset~\cite{hfa2024}, spanning 10 pitches from F3 to E$\flat$4 and covering varied harmonic patterns, including arpeggios, intervals, scales, and glissandi; each segment is 5.0\,s long. The second comprises 10 excerpts from Bergevin~et~al.~\cite{bergevin2020}, featuring four Tuvan singers (T1--T4) recorded in a sound booth. Both datasets were originally sampled at 96\,kHz and downsampled to 16\,kHz for synthesis, with mean segment durations of 5.0\,s (HFA) and $4.6 \pm 1.7$\,s (Bergevin).  

Copy-synthesis is particularly suitable when dealing with short recordings as it focuses on the specifics of individual frames. Each audio file is optimized independently from a single random initialization, without restarts, for up to 500 Adam iterations. For each recording, the reported metrics are computed at the iteration with the lowest training loss. All reported values are means $\pm$ standard deviations across recordings within each dataset. 

We evaluate log-spectral distance (LSD), spectral Pearson correlation (SpCorr: correlation between flattened log-magnitude spectrograms of target and synthesized signals), PESQ (Perceptual Evaluation of Speech Quality; ITU-T P.862, MOS 1--5\textcolor{black}{)~\cite{941023}}, CDPAM (Concatenated Deep Perceptual Audio Metric; learned distance metric, lower=better\textcolor{black}{)~\cite{manocha2021}}, ViSQOL (Virtual Speech Quality Objective Listener; spectro-temporal similarity, MOS-like 1--5\textcolor{black}{)~\cite{chinen2020visqol}}, plus mean subjective ratings from MUSHRA tests\textcolor{black}{~\cite{mushra2014}} for overall quality (Q1) and harmonic similarity (Q2); as well as the energy ratio overtone salience, and harmonic prominence ratio (HPR\textcolor{black}{, defined as the per-frame ratio of peak to mean magnitude inside the 1--3\,kHz band, averaged over time}) in the overtone region. \textcolor{black}{The MUSHRA-inspired listening test was run with 23 listeners (3 expert, 2 trained, 18 novice) on 6 segments (3 per dataset) in a blind A/B/C rating with randomized order.}

\begin{table*}[t]
\centering
\caption{Copy-synthesis results on 20 segments of Tuvan sygyt from two independent datasets.
Objective metrics: mean $\pm$ std across $N$ segments. Best in \textbf{bold}.
LSD and SpCorr are signal-level metrics; PESQ (ITU-T P.862.2; wideband MOS-LQO), CDPAM (learned perceptual distance, lower$=$closer), and ViSQOL (audio-mode MOS-LQO, 1--5) are objective perceptual metrics. Q1 (overall quality) and Q2 (harmonic similarity) are subjective scores (mean $\pm$ 95\% CI, 0--100 scale, 23~listeners $\times$ 3~segments per dataset). \textcolor{black}{DOFs are total trainable parameters per audio segment (per-frame tract coefficients plus global source parameters). The per-frame counts (13 articulator, 70 B-spline) appear in Sec.~\ref{sec:parameterization}.}}
\label{tab:main}
\footnotesize
\begin{tabular*}{\textwidth}{@{\extracolsep{\fill}}l r c c c c c c c c}
\toprule
 & DOFs & $N$ & LSD (dB)$\downarrow$ & SpCorr$\uparrow$ & PESQ$\uparrow$ & CDPAM$\downarrow$ & ViSQOL$\uparrow$ & Q1$\uparrow$ & Q2$\uparrow$ \\
\midrule
\multicolumn{10}{l}{\textit{HFA~\cite{hfa2024} --- 1 singer, 10 pitches (F3--E$\flat$4)}} \\
Artic.\ chain & 19  & 10 & 13.84 $\pm$ 0.54 & 0.71 $\pm$ 0.04 & 1.17 $\pm$ 0.15 & 3.06 $\pm$ 3.34 & 2.81 $\pm$ 0.53 & 13.6 $\pm$ 3.7 & 25.3 $\pm$ 4.5 \\
DDSP          & ${\sim}$100k & 10 & 10.99 $\pm$ 0.54 & 0.82 $\pm$ 0.01 & 1.20 $\pm$ 0.13 & 3.56 $\pm$ 3.71 & \textbf{3.81 $\pm$ 0.21} & 42.4 $\pm$ 5.0 & 50.3 $\pm$ 4.9 \\
B-spline      & 86  & 10 & \textbf{9.64 $\pm$ 0.29} & \textbf{0.86 $\pm$ 0.02} & \textbf{1.37 $\pm$ 0.34} & \textbf{2.36 $\pm$ 3.07} & 3.60 $\pm$ 0.22 & \textbf{44.8 $\pm$ 5.5} & \textbf{52.9 $\pm$ 5.8} \\
\midrule
\multicolumn{10}{l}{\textit{Bergevin~\cite{bergevin2020} --- 4 singers (T1--T4)}} \\
Artic.\ chain & 19  & 10 & 14.53 $\pm$ 0.93 & 0.66 $\pm$ 0.05 & 1.10 $\pm$ 0.05 & 0.73 $\pm$ 0.41 & 2.88 $\pm$ 0.21 & 11.6 $\pm$ 3.6 & 15.7 $\pm$ 3.9 \\
DDSP          & ${\sim}$100k & 10 & 10.71 $\pm$ 0.72 & 0.83 $\pm$ 0.03 & 1.26 $\pm$ 0.13 & 0.77 $\pm$ 0.55 & 3.69 $\pm$ 0.33 & 35.4 $\pm$ 5.5 & 45.1 $\pm$ 5.4 \\
B-spline      & 86  & 10 & \textbf{9.04 $\pm$ 0.46} & \textbf{0.88 $\pm$ 0.01} & \textbf{1.58 $\pm$ 0.35} & \textbf{0.60 $\pm$ 0.34} & \textbf{3.85 $\pm$ 0.23} & \textbf{38.5 $\pm$ 5.3} & \textbf{46.3 $\pm$ 5.6} \\
\bottomrule
\end{tabular*}
\end{table*}

\subsection{Pitch and Overtone Detection}

The dual-$f_0$ extraction pipeline described in Section~\ref{sec:f0_detection} successfully detects diphonic content in all 20 recordings. Across voiced frames, 99\% are classified as diphonic (enhancement $> 6$\,dB), with a mean enhancement of 37.6\,dB. The detected overtone trajectories correspond to harmonic numbers ranging from $H_2$ to $H_{10}$, consistent with the expected \emph{sygyt} overtone range~\cite{bergevin2020}. These extracted primary-$f_0$ and overtone trajectories are held fixed across all synthesis conditions.

\subsection{Overall Quality of Copies}

Table~\ref{tab:main} summarizes copy-synthesis performance across all 20 recordings for three methods: articulator chain, DDSP, and B-spline waveguide. All metrics favor the B-spline, substantially outperforming the articulator chain and, to a lesser extent, the baseline DDSP.

\begin{figure}[!b]
\centering
\includegraphics[width=\columnwidth]{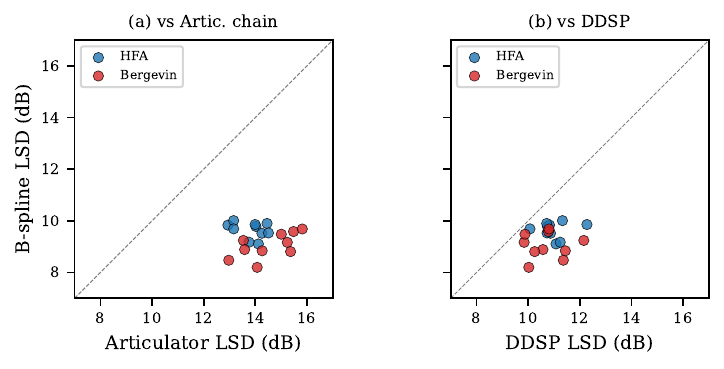}
\caption{Per-segment LSD comparison. Each point corresponds to one of the 20 recordings. Points below the diagonal indicate B-spline superiority. All 20 segments favor B-spline.}
\label{fig:segment_scatter}
\end{figure}

LSD shows B-spline reduces error by 30\% on HFA recordings and 38\% on Bergevin recordings, relative to articulator chain. 
B-spline also achieves lower LSD than DDSP on both datasets, with the lowest values on all 20 individual segments ---see Fig.~\ref{fig:segment_scatter} confronting LSD measures for B-spline with the articulator chain and DDSP methods on the HFA (blue) and Bergevin (red) datasets. The results yield large effect sizes (standardized differences in group means, with Cohen's $d=2.1$ for HFA and $d=1.9$ for Bergevin) with strong practical significance.

Objective perceptual metrics corroborate these findings on PESQ and CDPAM. On ViSQOL B-spline and DDSP largely outperform the articulator chain. 
\textcolor{black}{Subjectively, B-spline scored highest for overall quality and overtone similarity across datasets, followed by DDSP and the articulator chain. These subjective results corroborate the objective perceptual metrics, confirming that the spectral advantages of the B-spline waveguide are perceptible to human listeners.}

Friedman tests confirm significant method effects ($p<10^{-4}$). B-spline and DDSP significantly outperform articulator chain (Wilcoxon $p<0.001$, Cliff's $\delta>0.7$). \textcolor{black}{B-spline and DDSP do not differ significantly under MUSHRA ($p=0.19$); B-spline does, however, achieve consistently lower objective error (LSD, cepstral peak) than both baselines.}

\begin{figure}[t]
\centering
\includegraphics[width=\columnwidth]{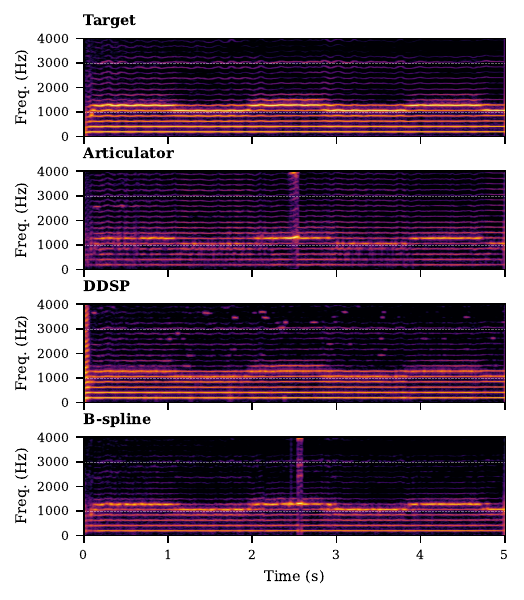}
\caption{Spectrogram comparison (0--4\,kHz) of the same recording ($h_5 \leftrightarrow h_6$). Rows: target, articulator chain, DDSP, and B-spline. Dashed lines indicate the 1--3\,kHz overtone region. \textcolor{black}{The short broadband transient at the start of the DDSP row is an initialization artifact of the harmonic-plus-noise oscillator bank and is excluded from all reported metrics.}}
\label{fig:spectrograms}
\end{figure}

\begin{table}[!t]
\centering
\caption{\textcolor{black}{Overtone-region errors $|\Delta|$ to the per-segment target (mean $\pm$ 1\,std across segments, lower is better). $|\Delta\mathrm{SpCorr}_\text{OT}| = 1-\mathrm{SpCorr}_\text{OT}$; remaining columns are $|x_\text{synth}-x_\text{target}|$ for the Bergevin energy ratio $eR$ (1--2/0--8\,kHz), overtone salience $S_\text{ot}$ (dB) and HPR. Best in \textbf{bold}.}}
\label{tab:overtone_region}
\footnotesize
\setlength{\tabcolsep}{3pt}
\resizebox{\columnwidth}{!}{%
\begin{tabular}{@{}llcccc@{}}
\toprule
Dataset & Method & $|\Delta\mathrm{SpCorr}_\text{OT}|\!\downarrow$ & $|\Delta eR|\!\downarrow$ & $|\Delta S_\text{ot}|$ (dB)$\!\downarrow$ & $|\Delta\mathrm{HPR}|\!\downarrow$ \\
\midrule
\multirow{3}{*}{HFA}    & Artic.\ chain & 0.21 $\pm$ 0.05          & 0.36 $\pm$ 0.12          & 1.83 $\pm$ 1.07          & 14.16 $\pm$ 7.77 \\
                        & DDSP          & \textbf{0.10 $\pm$ 0.02} & 0.46 $\pm$ 0.11          & 1.68 $\pm$ 0.84          & 15.69 $\pm$ 3.39 \\
                        & B-spline      & 0.12 $\pm$ 0.03          & \textbf{0.12 $\pm$ 0.05} & \textbf{0.88 $\pm$ 0.66} & \textbf{5.83 $\pm$ 2.31} \\
\midrule
\multirow{3}{*}{Berg.}  & Artic.\ chain & 0.48 $\pm$ 0.21          & 0.36 $\pm$ 0.16          & 2.53 $\pm$ 1.28          & 7.52 $\pm$ 3.98 \\
                        & DDSP          & \textbf{0.14 $\pm$ 0.08} & 0.23 $\pm$ 0.20          & 1.73 $\pm$ 1.07          & 9.16 $\pm$ 4.62 \\
                        & B-spline      & 0.18 $\pm$ 0.12          & \textbf{0.11 $\pm$ 0.08} & \textbf{1.04 $\pm$ 0.73} & \textbf{5.13 $\pm$ 1.47} \\
\midrule
\multirow{3}{*}{All}    & Artic.\ chain & 0.34 $\pm$ 0.21          & 0.36 $\pm$ 0.14          & 2.18 $\pm$ 1.23          & 10.84 $\pm$ 7.01 \\
                        & DDSP          & \textbf{0.12 $\pm$ 0.06} & 0.35 $\pm$ 0.20          & 1.70 $\pm$ 0.97          & 12.43 $\pm$ 5.20 \\
                        & B-spline      & 0.15 $\pm$ 0.09          & \textbf{0.12 $\pm$ 0.07} & \textbf{0.96 $\pm$ 0.70} & \textbf{5.48 $\pm$ 1.97} \\
\bottomrule
\end{tabular}}
\end{table}


\subsection{Acoustic Analysis}
\label{subsec:overtono}

\noindent\textbf{Overtone analysis}. \textcolor{black}{Table~\ref{tab:overtone_region} reports per-segment absolute errors to the target in the 1--3\,kHz overtone band. B-spline yields the lowest error in $eR$, $S_\text{ot}$ and HPR across both datasets, with overall errors of $|\Delta eR|=0.12\!\pm\!0.07$, $|\Delta S_\text{ot}|=0.96\!\pm\!0.70$\,dB and $|\Delta\mathrm{HPR}|=5.5\!\pm\!2.0$, clearly below articulator ($0.36$, $2.18$\,dB, $10.8$) and DDSP ($0.35$, $1.70$\,dB, $12.4$). DDSP edges B-spline in $|\Delta\mathrm{SpCorr}_\text{OT}|$ ($0.12$ vs.\ $0.15$ overall), consistent with its direct per-harmonic amplitude control; Fig.~\ref{fig:spectrograms} visualises the corresponding $h_5\!\leftrightarrow\!h_6$ overtone ridge. DDSP thus matches the local spectrum better, but B-spline preserves the global prominence and spectral balance characteristic of \emph{sygyt}.}



\begin{figure}[!b]
\centering
\includegraphics[width=\columnwidth]{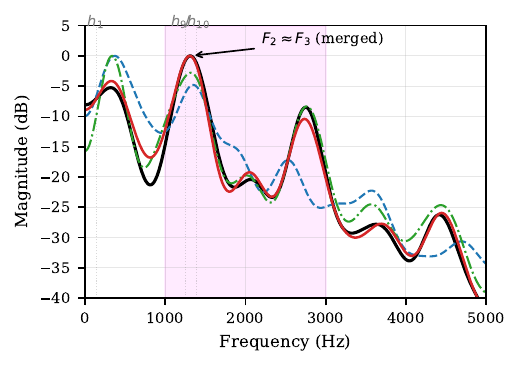}
\caption{Cepstral spectral envelopes ($N_\text{ceps}=30$) for the Bergevin T1 segment at $t=1.9$\,s (midpoint). \textcolor{black}{Curves: target (black, solid), articulator chain (blue, dashed), DDSP (green, dash-dot), B-spline (red, solid).}}
\label{fig:formant_envelope}
\end{figure}


\noindent\textbf{Formant analysis}. Cepstral envelope analysis shows that B-spline more accurately reproduces the target formant-merging pattern, a hallmark of \emph{sygyt} production in which $F_2$ and $F_3$ converge to amplify a single overtone~\cite{bergevin2020} (example illustrated in Fig.~\ref{fig:formant_envelope}). Its $C^2$-continuous tract profiles place the merged peak closer to the target than the articulator chain, whose coarse tongue-position control limits its ability to form the narrow constrictions required for precise formant placement. This formant-merging behavior is not explicitly imposed by the loss function. Rather, it emerges from optimization and is consistent with the formant-focusing account described by Bergevin~\textit{et~al.}~\cite{bergevin2020} (see Section~\ref{subsec:formant}). \textcolor{black}{Quantitatively, the cepstral peak location of the B-spline envelope tracks the target peak with the lowest error of the three methods (see below), which we read as evidence of more accurate formant merging.}

Across all 20 recordings, B-spline achieves a mean formant-peak error of 28\,Hz \textcolor{black}{(measured as the absolute deviation between the cepstral-envelope peak locations of target and synthesis in the 1--3\,kHz band)} relative to the target cepstral peak location in the 1--3\,kHz overtone region, compared with 222\,Hz for the articulator chain and 120\,Hz for DDSP. Its formant peak prominence (12.0\,dB) also closely matches that of the target recordings (12.4\,dB), suggesting that the optimizer captures the location of the merged peak and part of its spectral prominence.

\begin{figure}[!t]
\centering
\includegraphics[width=\columnwidth]{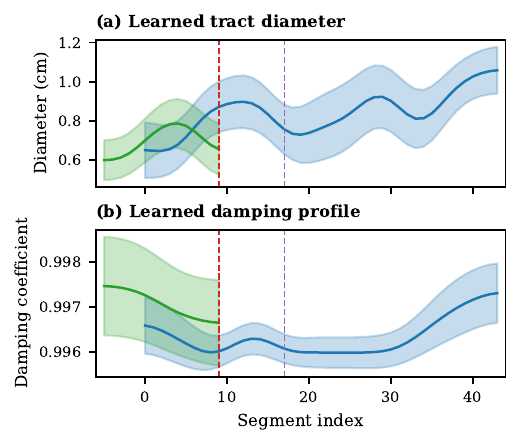}
\caption{Learned B-spline profiles for a representative segment (A3 interval): oral and sublingual diameters (top) and dampings (bottom) along the tract. \textcolor{black}{Solid lines: oral (blue), sublingual (green); dashed verticals: sublingual junction (red), velum (purple); shaded bands: $\pm 1$\,SD across segments.}}
\label{fig:tract_damping}
\end{figure}

\subsection{Learning Process for Tract Parameterizations}

We examine learned parameters from the physically-grounded baselines, B-spline and articulator chain, excluding DDSP due to its signal-domain harmonic parameterization, which does not yield tract configurations amenable to acoustic interpretation. \textcolor{black}{The articulator chain exposes a single global damping coefficient rather than a per-segment damping profile, so equivalent damping panels cannot be drawn for it; we focus on the B-spline profiles below.}


\noindent\textbf{Learned Tract Shape and Damping}.
The B-spline parameterization yields tract profiles for diameter (top) and damping (bottom), such as the one in Fig.~\ref{fig:tract_damping}. These foster a physically meaningful acoustic interpretation of the oral \textcolor{black}{(blue)} and sublingual (green) waveguides. Junctions are marked with dashed lines. The oral tract exhibits pronounced constriction in segment 9 (sublingual junction), consistent with the position of the posterior tongue observed in MRI studies of sygyt production~\cite{adachi1999}. This configuration emerges end-to-end from gradient descent without explicit anatomical constraints. This pattern holds in 16 of 20 optimized segments (80\%), covering the overtones $h_5$ to $h_{10}$.

The damping profile ($d\in[0.99,0.9999]$, bottom part of Fig.~\ref{fig:tract_damping}) exhibits complementary spatial structure. Lower values ($d\to0.99$) near the sublingual junction increase energy dissipation, effectively isolating the proximal tract from the anterior oral cavity. Higher values ($d\to0.9999$) within the anterior resonance cavity sustain narrowband, high-$Q$ resonances (i.e., high quality factor $Q=\textcolor{black}{f_c}/\Delta f$, for \textcolor{black}{resonance centre frequency $f_c$ (not the voice $f_0$)} and bandwidth $\Delta f$) characteristic of the 1--3 kHz overtone region. The sublingual branch maintains uniformly low damping, consistent with reduced acoustic coupling during overtone production.

Together, these profiles suggest a coherent mechanism for \emph{sygyt} spectral focusing: tract constriction aligns relevant formants, junction damping limits energy backscatter, and the anterior cavity supports overtone resonance---all without introducing physiology-specific loss terms into the training losses.

\begin{figure}[!t]
\centering
\includegraphics[width=.8\columnwidth]{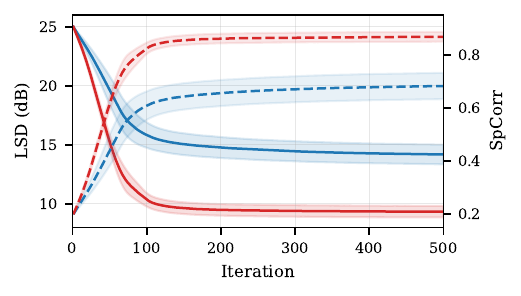}
\caption{Convergence analysis on both physical-model synthesis, articulator chain (blue) and B-spline (red), averaged across all segments (shading: $\pm 1\,\sigma$). Solid lines show LSD (left axis, dB~$\downarrow$); dashed lines show SpCorr (right axis, $\uparrow$).}
\label{fig:convergence}
\end{figure}

\noindent\textbf{Convergence}. 
Despite substantially more learnable parameters (70 vs.~13 DOFs/frame), B-spline optimization converges at rates comparable to the articulator chain baseline (Fig.~\ref{fig:convergence}). Through the initial 300 iterations, B-spline maintains lower LSD and higher spectral correlation throughout the trajectory. These dynamics indicate that the flexible tract representation facilitates discovery of acoustically interpretable configurations within equivalent computational budgets. \textcolor{black}{Optimization is offline ($\sim$30\,min per 5\,s segment on CPU, RTF $\approx 360\times$); real-time use is out of scope.} 




\section{Ablation Study}\label{sec:ablation}


\begin{table}[!b]
\centering
\caption{$2 \times 2$ factorial ablation (all 20 segments, 500 iterations). Sub.\ = sublingual second source; Damp.\ = per-segment damping. Values are mean $\pm$ std across recordings.}
\label{tab:ablation_factorial}
\small
\setlength{\tabcolsep}{3.5pt}
\resizebox{\columnwidth}{!}{%
\begin{tabular}{@{}lccccc@{}}
\toprule
Condition & Sub. & Damp. & LSD (dB)$\downarrow$ & SpCorr$\uparrow$ & SpCorr$_\text{OT}$$\uparrow$ \\
\midrule
Full           & \checkmark & \checkmark & \textbf{9.34$\pm$0.49} & \textbf{0.86$\pm$0.02} & \textbf{0.85$\pm$0.09} \\
No sublingual  & --         & \checkmark & 10.32$\pm$0.56 & 0.82$\pm$0.03 & 0.81$\pm$0.13 \\
No damping     & \checkmark & --         & 9.48$\pm$0.48 & 0.86$\pm$0.02 & 0.85$\pm$0.09 \\
Minimal        & --         & --         & 10.45$\pm$0.61 & 0.82$\pm$0.03 & 0.81$\pm$0.14 \\
\bottomrule
\end{tabular}%
}
\end{table}

To disentangle the contributions of the sublingual second source and spatially variable damping, we perform a $2 \times 2$ factorial ablation of the B-spline condition across all 20 recordings (Table~\ref{tab:ablation_factorial}). \textcolor{black}{This ablation targets architectural components; a separate ablation of the loss terms $\mathcal{L}_\text{mel}$, $\mathcal{L}_\text{harm}$, $\mathcal{L}_\text{energy}$, $\mathcal{L}_\text{ot}$ is left to future work.} Removing the sublingual source increases LSD more strongly (+1.0\,dB) than removing spatially variable damping (+0.1\,dB), indicating that the secondary excitation mechanism is the stronger contributor under this experimental setup. Spatially variable damping still provides a consistent but smaller improvement. The minimal condition, with neither sublingual source nor spatially variable damping, performs worst overall. The combined pattern also suggests only a limited interaction between both components, as each provides a largely separable benefit.

\section{Discussion}\label{sec:discussion}

Our model includes a sublingual tube that, while not anatomically human, provides a controllable physical system for studying the acoustics of diphonic singing. \textcolor{black}{We treat the sublingual branch as an \emph{acoustic abstraction}, not as a literal anatomical claim: it is one way to introduce the additional spectral degree of freedom that Bergevin et~al.~\cite{bergevin2020} attribute to precise linear filtering of a single source.} The $2 \times 2$ factorial ablation (Section~\ref{sec:ablation}) indicates that the sublingual source is the stronger contributor under the present experimental setup, with a substantially larger effect on LSD than spatially variable damping. This suggests that the secondary excitation mechanism has a larger impact on the reported reconstruction metrics than fine-grained Q-factor control alone. Spatially variable damping provides a slight secondary improvement, suggesting that tract shaping and formant focusing remain relevant even when the secondary source dominates the metric gains. This is a relevant result that we wanted to track for future work.

The DDSP comparison tests whether physical modeling remains useful relative to an unconstrained signal-space baseline. DDSP offers direct per-harmonic amplitude control without any tract-based inductive bias, yet the waveguide achieves lower reconstruction error overall. This suggests that explicit acoustic structure is beneficial in this copy-synthesis setting. More specifically, the waveguide constrains the solution to spectra that are realizable through tube resonances and source--filter interaction, while still allowing the optimizer to discover tract configurations that are interpretable in acoustic terms. We therefore view the advantage of the physical model not as proof of a specific physiological mechanism, but as evidence that a structured waveguide representation is a useful bias for overtone-singing reconstruction.

\subsection{Limitations}

Six key limitations for \emph{sygyt} copy-synthesis warrant discussion: 
(1) moderate absolute LSD (9--15\,dB) reflects the intrinsic difficulty of matching overtone-singing spectra with a simplified physical model, but relative 30--38\% gains over baseline show efficacy; 
(2) sublingual source is only an approximation to complex aeroacoustic interactions in diphonic singing (+1.0\,dB ablation gain), where some secondary pitches may arise without distinct vibrating structures; 
(3) while $N=20$ curated segments from two independent datasets (5 singers, 10 pitches) effectively demonstrate proof-of-concept across varied performers and pitches—with low per-recording standard deviation ($\sigma_{LSD} < 0.93$ dB)— and prioritize pattern diversity (arpeggios, glissandi, steady tones) over volume, future work should extend our analysis; the present scale prioritizes diversity over volume, aligning with the niche nature of high-quality \emph{sygyt} data;
(4) \emph{sygyt}-specific scope excludes other \emph{khoomei} styles like \emph{kargyraa} or uncurated field recordings for broader robustness;
(5) fixed pitch extraction limits glissandi but achieves 86\% correlation, fixable via joint optimization;
plus uncharacterized real-time potential post-optimization\textcolor{black}{; (6) DDSP uses a neural encoder vs.\ our per-frame parameters (both run to convergence with the same loss), and porting our sublingual mechanism to DDSP requires non-trivial architectural changes -- encoder-matched DDSP and an SVS extension with unfrozen $f_0$/overtone are left to future work}.

\section{Conclusion}

We have presented copy-synthesis methods for \emph{sygyt}-style singing, based on differentiable DSP and a differentiable Kelly--Lochbaum waveguide synthesizer extended for diphonic singing. The model combines a sublingual second glottal source with a cubic B-spline spatial parameterization. The B-spline basis uses a compact set of control points for oral-tract diameter and damping, yielding smooth tract profiles and enabling spatially structured learnable damping for fine-grained control of resonant behavior.

\sloppy
Copy-synthesis experiments from two independent datasets (5 singers, 10 pitches) show that the B-spline parameterization reduces LSD by 30--38\% relative to the articulator-chain baseline, while also outperforming a DDSP baseline in overall LSD and global spectral correlation. The gain is concentrated in the 1--3\,kHz overtone region, where the model achieves an average improvement of $4.7 \pm 1.8$\,dB over the articulator chain and reproduces the target merged-formant peak with a mean error of 28\,Hz. The learned tract profiles also admit a plausible acoustic interpretation, with constrictions and localized damping structure consistent with overtone-focused spectral shaping. A $2 \times 2$ factorial ablation further shows that the sublingual second source is the stronger contributor under the present setup, while spatially structured damping provides a smaller complementary benefit.


\section{Acknowledgments and Supp. Materials}

The authors would like to thank Christopher Bergevin for their valuable contribution to this work. Audio examples and supplementary materials are available on the companion website.\footnote{\textcolor{black}{\url{https://mateocamara.com/khoomei-supp-materials}}}
\textcolor{black}{This work was supported by the Ministry of Economy and Competitiveness of Spain under grant PID2021-128469OB-I00, and «Ayuda Económica Para Personal Investigador Postdoctoral 2025» of «Fundación Santander».}

\begingroup
\renewcommand{\baselinestretch}{0.95}
\scriptsize
\bibliographystyle{IEEEtranDAFx}
\bibliography{DAFx26_tmpl} 
\endgroup

\end{document}